# Macroscopic violation of special relativity

G. Nimtz[1, 2] and A. A. Stahlhofen[1]

**Feynman, one of the founders of Quantum Electrodynamics (QED) introduced virtual particles in his diagrams as intermediate states of an interaction process. Such virtual particles are not observable. However, from the theoretical point of view, they represent necessary intermediate states between observable real states. Such virtual particles were introduced for describing the interaction process between an electron and a positron and for much more complicated interaction processes. Other candidates for virtual photons are evanescent modes in optics. Evanescent modes have a purely imaginary wave number and represent the mathematical analogy of the tunneling solutions of the Schrödinger equation. Evanescent modes are present in the forbidden frequency bands of a photonic lattice and in undersized wave guides for instance. The most prominent example of the occurrence of evanescent modes is frustrated total internal reflection (FTIR) at double prisms. In 1949 Sommerfeld[1] pointed out that this optical phenomenon represents the analogy of quantum mechanical tunneling. The evanescent modes (i.e. tunneling) lie outside the bounds of the special theory of relativity.**

**We demonstrate the quantum mechanical behavior of evanescent modes with digital microwave signals at a macroscopic scale of the order of a meter and show that evanescent modes are well described by virtual photons as predicted by former QED calculations.**

Evanescent modes or photonic tunneling like tunneling solutions of the Schrödinger equation have a purely imaginary wave number. This means that they do not experience a phase shift in traversing space.

The time delay $\tau$ of a wave packet is given by

$$\tau = d\varphi / d\omega \quad (1)$$

where $\varphi$ is the phase shift of the mode or of a particle and $\omega$ is the angular frequency.

In general, $\varphi$ is given by the real part of the wave number $k$ times the distance $x$. In the case of evanescent modes and tunneling solutions the real part of $k$ is zero. In view of Eq. (1), propagation across barriers appears to take place in zero time. In the case of particle tunneling Eq. 1 is replaced by the corresponding derivative of the S-matrix.

Despite the fact that the semiconductor tunnel or Esaki diode has been used since 1962., the particle barrier penetration time has not yet been determined due to parasitic time consuming electronic interaction effects in a semiconductor. Around 1990 the mathematical analogy between the Schrödinger and the Helmholtz equations inspired microwave and optical tunneling experiments to obtain empirical data on the tunneling time. The experiments with evanescent modes revealed superluminal energy and signal velocities [2, 3, 4, 5].

Several QED and QM calculations predicted that both evanescent modes and tunneling particles appear to propagate in zero time[6, 7, 8, 9, 10]. The only time delay arises as a scattering time at the barrier front, which has been shown to be a universal quantity of the order of magnitude of the time of the inverse center frequency of the tunneling wave packet[11, 12]. We now introduce an experiment with evanescent modes, which has proved the QED calculations.

Experiments with evanescent modes were carried out with undersized wave guides and photonic band gap material[11]. The virtual nature of evanescent modes becomes most obvious, however, in the time honoured example of frustrated internal reflection with double prisms.

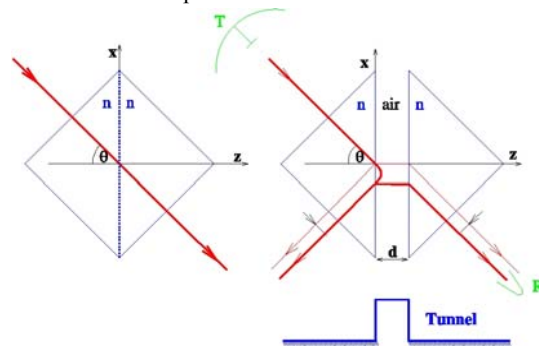

**Figure 1: Frustrated total internal reflection (FTIR) at double prisms.** The beam crossing the closed and the open prisms. The latter case with the gap corresponds to 1d-quantum mechanical tunneling [1]. The reflected/ transmitted beams both experience a longitudinal shift in the plane of incidence coined Goos-Hänchen shift as sketched in the figure.

The experimental set-up is illustrated in Fig.1. We have investigated double prisms of Perspex with a refractive index $n = 1.6$ with microwaves at a frequency of 9.15 *GHz*, i.e. at a wavelength of 32.8 *mm*. The sides of the right triangle prisms are 0.4 x 0.4 $m^2$, which is of a macroscopic dimension for a quantum mechanical experiment. The experiment was carried out with a symmetrical beam path as sketched in Fig. 1. The beam has a perpendicular incidence at the first prism and is reflected at the Perspex/ air boundary under an angle of 45°, which is above the critical angle of total reflection. (The critical angle for the Perspex prism is 38.7°.) The dish antennas had diameters of 350 *mm;* the receiver antenna was movable parallel to the prism's surfaces. The microwave polarization was TM, with the electric field in the plane of incidence. The measured Goos-Hänchen shift in this experiment is of the order of a wavelength[13].

The experimental result is that the reflected and the transmitted (tunneled) signal are received by the detectors at the same time. The measured time delay in both reflection and transmission of the digital pulse is about 100 ps. In this case of FTIR, this measured time delay corresponds to the Goos-Hänchen shift along the barrier boundary, i.e. along the prism-air boundary. The result is in agreement with the universal tunneling time cited above [11, 12].

This universal tunneling time seems to hold even for sound waves (i.e. phonons) as measured by Yang et al. at 1 MHz and by Robertson et al. at 1 kHz in a sound tunneling experimental set-up[14, 15]. Presumably, the virtual behavior of photons discussed here applies for all fields with wave solutions having purely imaginary wave numbers.

Evanescent modes have a purely imaginary wave number k. This violates the Einstein relation

$$E^2 = (\hbar k c)^2. \qquad (2)$$

In consequence of the phase time approach of Eq.1, the zero phase shift of spreading evanescent modes implies that barriers are crossed in zero time. This theoretical expectation is confirmed in the experiment shown above (cf. Fig. 1): The paths of the reflected and of the transmitted beams are equal except that the transmitted one has in addition crossed the evanescent region of length d.

Another property not compatible with classical physics is the fact that an evanescent mode is not observable. This follows from the uncertainty relation[10]. In order to observe an evanescent mode in the exponential tail of the evanescent region it must be localized within a distance of order of $\Delta x \approx ik = \kappa$. The uncertainty relation implies that

$$\Delta p > \hbar / \Delta x \approx \hbar \kappa = (k_o^2 (n_2^2 \sin^2(\varphi) - n_1^2))^{1/2}, \qquad (3)$$

where $k_o = \omega / c$ and $n_1$ and $n_2$ are the refractive indices of the air and of the prisms, respectively. The evanescent mode can thus be located in the nonclassical region only if a refractive index $\Delta n$ is added with $\Delta n = (k_o^2 (n_2^2 \sin^2(\varphi) - n_1^2))^{1/2}$ thus raising the evanescent mode into the classically allowed region.

All three properties - the violation of the Einstein energy relation, the zero time spreading, and the non observability of evanescent modes - can be explained by identifying evanescent modes with virtual photons as predicted by several authors, see for instance references [6, 7, 8, 9, 10]. The corresponding Feynman diagram is sketched in Fig. 2. Tunneling and evanescent modes are properly described by quantum mechanics.

We are used to virtual particles in microscopic interaction processes, but demonstrated here an example in the macroscopic range of a meter.

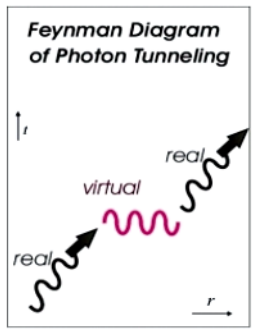

**Figure 2: Feynman diagram of evanescent mode or photonic tunneling.** Tunneling and evanescent modes are represented by a virtual photon.